\documentstyle[12pt]{article}

\def\thebibliography#1{\bigskip\section*{\centering
References\\}\bigskip\list
  {[\arabic{enumi}]}{\settowidth\labelwidth{#1}\leftmargin\labelwidth
    \advance\leftmargin\labelsep
    \usecounter{enumi}}
    \def\newblock{\hskip .11em plus .33em minus .07em}
    \sloppy\clubpenalty4000\widowpenalty4000
    \sfcode`\.=1000\relax}

\def\op#1{\mathop{\fam0 #1}\limits}
\newcommand{\Ker}{{\rm Ker\,}}
\newcommand{\nm}[1]{\mid {#1}\mid}
\newcommand{\beq}{\begin{equation}}
\newcommand{\eeq}{\end{equation}}
\newcommand{\ben}{\begin{eqnarray}}
\newcommand{\een}{\end{eqnarray}}
\newcommand{\be}{\begin{eqnarray*}}
\newcommand{\ee}{\end{eqnarray*}}
\newcommand{\bea}{\begin{eqalph}}
\newcommand{\eea}{\end{eqalph}}

\newcommand{\cL}{{\cal L}}

\newcommand{\cD}{{\cal D}}

\newcommand{\R}{{\bf R}}
\newcommand{\C}{{\bf C}}
\newcommand{\al}{\alpha}
\newcommand{\bt}{\beta}
\newcommand{\dl}{\delta}
\newcommand{\la}{\lambda}

\newcommand{\f}{\phi}

\newcommand{\om}{\omega}

\newcommand{\m}{\mu}
\newcommand{\n}{\nu}
\newcommand{\g}{\gamma}
\newcommand{\G}{\Gamma}

\newcommand{\ve}{\varepsilon}

\newcommand{\si}{\sigma}
\newcommand{\Si}{\Sigma}

\newcommand{\wt}{\widetilde}
\newcommand{\wh}{\widehat}
\newcommand{\ol}{\overline}
\newcommand{\dr}{\partial}

\newcommand{\ot}{\otimes}
\newcommand{\ap}{\approx}

\newenvironment{eqalph}{\stepcounter{equation}
\setcounter{equationa}{\value{equation}}
\setcounter{equation}{0}

\begin{eqnarray}}{\end{eqnarray}\setcounter{equation}{\value{equationa}}}

\begin{document}
\hbox{}

\begin{center}

{\Large \bf BACKGROUND GEOMETRY

IN GAUGE GRAVITATION THEORY}
\bigskip

{\sc Gennadi Sardanashvily}
\medskip

Department of Theoretical Physics, Moscow State University,

117234 Moscow, Russia

E-mail: sard@grav.phys.msu.su
\end{center}

\begin{abstract}
Dirac fermion fields are responsible for spontaneous symmetry breaking
in gauge gravitation theory because the spin structure associated with
a tetrad field is not preserved under general covariant transformations.
Two solutions of this problem can be suggested. (i) There exists the universal
spin structure $S\to X$ such that any spin structure $S^h\to X$
associated with a tetrad field $h$ is a subbundle of the bundle $S\to X$.
In this model, gravitational fields correspond to different tetrad
(or metric) fields. (ii) A background tetrad field $h$ and the associated
spin structure $S^h$ are fixed, while gravitational fields are
identified with additional tensor fields $q^\la{}_\m$ describing deviations 
$\wt h^\la_a=q^\la{}_\m h^\m_a$ of $h$. One can think of $\wt h$ as
being effective tetrad fields. We show that there exist gauge
transformations which keep the background tetrad field $h$ and act on
the effective fields by the general covariant transformation law.
We come to Logunov's Relativistic Theory of Gravity generalized
to dynamic connections and fermion fields. 
\end{abstract}

\section{Introduction}

Existence of Dirac fermion fields implies that, if a world manifold $X$
is non-compact in order to satisfy causility conditions, it is
parallelizable, that is, the tangent bundle $TX$ is trivial and the
principal bundle $LX$ of oriented frames in $TX$ admits a global
section \cite{ger}.

Dirac spinors are defined as follows \cite{law,cra}.
Let $M$ be the Minkowski space with the metric 
\be
\eta ={\rm diag}(1,-1,-1,-1),
\ee
written with respect to a basis $\{e^a\}$.
By $\C_{1,3}$ is meant the complex Clifford algebra generated by $M$. 
This is isomorphic to the real Clifford algebra $\R_{2,3}$ over $\R^5$.
Its subalgebra generated by $M\subset\R^5$ is the real Clifford algebra
$\R_{1,3}$. A Dirac spinor space $V_s$ is defined as a minimal left
ideal of $\C_{1,3}$ on which this algebra acts on the left.
We have the representation  
\beq
\g: M\otimes V_s \to V_s, \qquad \wh e^a=\g(e^a)=\g^a, \label{w01}
\eeq
of elements of the Minkowski space $M\subset\C_{1,3}$ by the Dirac
matrices on $V_s$. The explicit form of this representation depends on
the choice of the ideal $V_s$. Different ideals lead to equivalent
representations (\ref{w01}). The spinor space $V_s$ is provided with
the spinor metric 
\beq
a(v,v') =\frac12(v^+\g^0v' +{v'}^+\g^0v).\label{b3201}
\eeq

Let us consider morphisms preserving the representation (\ref{w01}).
By definition, the Clifford group $G_{1,3}$ consists of the invertible
elemets $l_s$ of the real Clifford algebra $\R_{1,3}$ such that the
inner automorphisms 
\beq
l_sel^{-1}_s = l(e), \qquad e\in M, \label{b3200}
\eeq
defined by these elements preserve the Minkowski space $M\subset
\R_{1,3}$. Since the action (\ref{b3200}) of the group $G_{1,3}$ on $M$
is not effective, one usually considers its spin subgroup 
\be
L_s={\rm Spin}^0(1,3)\simeq {\rm SL}(2,\C).
\ee
This is the two-fold universal covering group $z_L:L_s\to L$ of the
proper Lorentz group $L=SO^0(1,3)$. The group $L$ acts on $M$ by the
generators  
\beq
 L_{ab}{}^c{}_d= \eta_{ad}\dl^c_b- \eta_{bd}\dl^c_a. \label{b3278}
\eeq
The spin group $L_s$ acts on the spinor space $V_s$ by the generators
\beq
L_{ab}=\frac{1}{4}[\g_a,\g_b]. \label{b3213}
\eeq
Since, $L_{ab}^+\g^0=- \g^0L_{ab}$, the group $L_s$ preserves the
spinor metric (\ref{b3201}). 
The transformations (\ref{b3278}) and (\ref{b3213}) preserve the 
representation (\ref{w01}), that is,  
\be
\g (lM\otimes l_sV_s) = l_s\g (M\otimes V_s). 
\ee

A Dirac spin structure on a world manifold $X$ is said to be a pair 
$(P_s, z_s)$ of an $L_s$-principal bundle $P_s\to X$ and a principal 
bundle morphism of $P_s$ to the frame bundle $LX$ with the structure
group $GL_4=GL^+(4,{\bf R})$
\cite{law,avis,benn}. Owing to the group epimorphism $z_L$, every bundle
morphism $z_s$ factorizes through a bundle epimorphism of $P_s$ onto
a subbundle $L^hX\subset LX$ with the structure group $L$. Such a
subbundle $L^hX$ is called a Lorentz structure \cite{zul,gord}. It
exists since $X$ is parallelizable.

By virtue of the well-known theorem \cite{kob}, there is one-to-one
correspondence between the Lorentz subbundles of the frame bundle $LX$
and the global sections $h$ of the quotient bundle
\beq
\Si=LX/L\to X. \label{5.15}
\eeq
This is the two-fold covering of the bundle of pseudo-Riemannian
metrics in $TX$, and sections of $\Si$ are tetrad fields.
Let $P^h$ be the $L_s$-principal bundle covering $L^hX$ and
\beq
S^h=(P^h\times V_s)/L_s\label{510}
\eeq
the associated spinor bundle. Sections $s_h$ of $S^h$ describe Dirac
fermion fields in the presence of the tetrad field $h$.

Indeed, every tetrad field $h$ yields the structure 
\be
T^*X= (L^hX\times M)/L 
\ee
of the $L^hX$-associated bundle of Minkowski spaces on $T^*X$, and defines
the representation 
\ben
&& \g_h: T^*X\ot S^h=(P^h\times (M\ot V_s))/L_s\to (P^h\times
\g(M\ot V_s))/L_s=S^h,\nonumber \\
&& \g_h: T^*X\ni t^*=\dot x_\la dx^\la\mapsto \dot x_\la\wh dx^\la=\dot
x_\la
h^\la_a(x)\g^a, \label{L4}
\een
of covectors on $X$ by the Dirac matrices on elements of the spinor
bundle $S^h$. The crucial point is that different tetrad fields $h$
define non-equivalent representations (\ref{L4}) \cite{sar91,sar92}.
It follows that every Dirac fermion field must be considered in a pair
with a certain tetrad field.

Thus, we come to the well-known problem of describing fermion fields in
the presence of different gravitational fields and under 
general covariant transformations. Recall that general covariant
transformations are automorphisms of the frame bundle $LX$ which are
the canonical lift of diffeomorphisms of a world manifold $X$. They do
not preserve the Lorentz subbundles of $LX$. The following two solutions of
this problem can be suggested. 

(i) One can describe the total system of the pairs of spinor and tetrad
fields if any spinor bundle $S^h$ is represented as a subbundle of some
fibre bundle
$S\to X$ [11-14]. To construct $S$, let us consider
the two-fold universal covering group $\wt{GL_4}$ of the group $GL_4$
and the corresponding principal bundle $\wt{LX}$ covering the frame
bundle $LX$ [2,15-17]. Note that the group 
$\wt{GL_4}$ admits only infinite-dimensional spinor representations \cite{heh}.
At the same time, $\wt{LX}$ has the structure of the $L_s$-principal
bundle $\wt{LX}\to \Si$ \cite{may97,book}. Then let us consider the
associated spinor bundle 
\be
S=(\wt{LX}\times V_s)/L_s\to \Si
\ee
which is the composite bundle 
$S\to\Si\to X$.
We have the representation 
\ben
&&\g_\Si: (\Si\op\times_X T^*X)\op\ot_\Si S\to S,\nonumber \\
&&\g_\Si: dx^\la\mapsto \si^\la_a\g^a, \label{a1}
\een
of covectors on $X$ by the Dirac matrices. One can show that, for any
tetrad field $h$, the restriction of
$S\to\Si$ to $h(X)\subset \Si$ is a subbundle of $S\to X$ which is
isomorphic to the spinor bundle $S^h$, while the representation   
$\g_\Si$ (\ref{a1}) restricted to $h(X)$ is exactly the representation $\g_h$
(\ref{L4}). The bundle $\wt{LX}$ inherits general covariant
transformations of $LX$ \cite{dabr}. They, in turn, induce general
covariant transformations of $S$, which transform the subbundles  
$S^h\subset S$ to each other and preserve the representation (\ref{a1})
\cite{may97,book}. 

(ii) This work is devoted to a different model.  
A background tetrad field $h$ and the associated background
spin structure $S^h$ are fixed, while gravitational fields are
identified with the sections of the group bundle $Q\to X$ associated with $LX$
(in the spirit of
Logunov's Relativistic Gravitation Theory (RGT) \cite{log}). 
We will show that there exists an automorphism $\wt f_h$ of $LX$ over any 
diffeomorphism $f$ of $X$  which preserves the Lorentz subbundle
$L^hX\subset LX$.

\section{Gauge transformations}

With respect to the tangent holonomic frames $\{\dr_\m\}$, the frame
bundle $LX$ is equipped with the coordinates $(x^\la, p^\la{}_a)$ such
that general covariant transformations of $LX$ over diffeomorphisms
$f$ of $X$ take the form
\be
\wt f: (x^\la, p^\la{}_a) \mapsto (f^\la(x), \dr_\m f^\la(x)p^\m{}_a).
\ee
They induce general covariant transformations 
\be
\wt f: (p,v)\cdot GL_4\mapsto (\wt f(p),v)\cdot GL_4
\ee
of any $LX$-associated bundle
\be
Y=(LX\times V)/GL_4,
\ee
where the quotient is defined by identification of elements
$(p,v)$ and $(pg,g^{-1}v)$ for all $g\in GL_4$.

Given a tetrad field $h$, any general covariant transformation of the frame
bundle $LX$ can be written as the composition
$\wt f=\Phi\circ \wt f_h$  of its automorphism $\wt f_h$ over $f$ which
preserves $L^hX$ and some vertical automorphism 
\beq
\Phi: p\mapsto p\f(p), \qquad p\in LX, \label{a3}
\eeq
where $\f$ is a $GL_4$-valued equivariant function on $LX$, i.e., 
\be
\f(pg)=g^{-1}\f(p)g, \qquad g\in GL_4.
\ee
Since $X$ is parallelizable, the automorphism $\wt f_h$ exists. Indeed,
let $z^h$ be a global section of $L^hX$. Then, we put
\be
\wt f_h: L_xX\ni p=z^h(x)g\mapsto z^h(f(x))g\in L_{f(x)}X.
\ee
The automorphism $\wt f_h$ restricted to $L^hX$ induces an automorphism
of the principal bundle $P^h$ and the corresponding automorphism 
$\wt f_s$ of the spinor bundle $S^h$, which preserve the representation 
(\ref{L4}). 

Turn now to the vertical automorphism $\Phi$. Let us consider the group
bundle $Q\to X$ associated with $LX$. Its typical fibre is the group
$GL_4$ which acts on itself by the adjoint representation. Let 
$(x^\la, q^\la{}_\m)$ be coordinates on $Q$.
There exist the left and right canonical actions of 
$Q$ on any $LX$-associated bundle $Y$:
\be
&&\rho_{l,r}: Q\op\times_X Y\to Y, \\
&&\rho_l: ((p,g)\cdot GL_4, (p,v)\cdot GL_4) \mapsto (p,gv)\cdot
GL_4,\\ 
&& \rho_r: ((p,g)\cdot GL_4, (p,v)\cdot GL_4) \mapsto (p,g^{-1}v)\cdot
GL_4. 
\ee

Let $\Phi$ be the vertical automorphism (\ref{a3}) of $LX$. The corresponding
vertical automorphisms of an associated bundle
$Y$ and the group bundle $Q$ read
\be
&& \Phi: (p,v)\cdot GL_4\mapsto (p, \f(p)v)\cdot GL_4, \\
&& \Phi: (p,g)\cdot GL_4\mapsto (p,\f(p)g\f^{-1}(p))\cdot GL_4.
\ee
For any $\Phi$ (\ref{a3}), there exists the fibre-to-fibre morphism 
\be
\ol\Phi: (p,q)\cdot GL_4\mapsto (p,\f(p)q)\cdot GL_4 
\ee
of the group bundle $Q$ such that 
\ben
&& \rho_l(\ol\Phi(Q)\times Y)=\Phi(\rho_l (Q\times Y)),\label{a6}\\
&& \rho_r(\ol\Phi(Q)\times \Phi(Y))=\rho_r (Q\times Y).\label{a6a}
\een

For instance, if $Y=T^*X$, the expression (\ref{a6}) takes the
coordinate form
\be
&&\rho_r: (x^\la,q^\la{}_\m, \dot x_\m)
\mapsto (x^\la,\dot x_\la q^\la{}_\m), \\
&& \ol\Phi: (x^\la,q^\la{}_\m) \mapsto (x^\la,S^\la{}_\nu q^\nu{}_\m), \\
&& \rho_r (x^\la,S^\la{}_\nu q^\nu{}_\m,
\dot x_\al(S^{-1})^\al{}_\la)= (x^\la,\dot x_\la q^\la{}_\m).
\ee
Hence, we obtain the representation  
\ben
&& \g_Q: (Q\times T^*X)\op\ot_Q (Q\times S^h) \to (Q\times S^h), \nonumber
\\
&&\g_Q=\g_h\circ\rho_r: (q,t^*)\mapsto 
\dot x_\la q^\la{}_\m\wh dx^\m=\dot x_\la q^\la{}_\m h^\m_a\g^a, \label{a8}
\een
on elements of the spinor bundle $S^h$. 
Let $q_0$ be the canonical global section of the group bundle $Q\to X$
whose values are the unit elements of the fibres of $Q$. Then, the
representation $\g_Q$ (\ref{a8}) restricted to $q_0(X)$ comes to the
representation $\g_h$ (\ref{L4}).

Sections $q(x)$ of the group bundle $Q$ are dynamic variables of the
model under consideration. One can think of them as being tensor
gravitational fields of Logunov's RTG. There is the canonical morphism
\be
&&\rho_l: Q\op\times_X\Si \to \Si, \\
&&\rho_l: 
((p,g)\cdot GL_4,(p,\si)\cdot GL_4)\mapsto (p,g\si)\cdot GL_4, 
\qquad p\in LX,\\
&&\rho_l: (x^\la,q^\la{}_\m, \si^\m_a)\mapsto 
(x^\la,q^\la{}_\m \si^\m_a). 
\ee
This morphism restricted to $h(X)\subset \Si$ takes the form
\ben
&&\rho_h: Q\to \Si, \nonumber \\
&&\rho_h: ((p,g)\cdot L,(p,\si_0)\cdot L)\to (p,g\si_0)\cdot L, 
\qquad p\in L^hX, \label{a11}\\
&&\rho_h: (x^\la,q^\la{}_\m)\mapsto (x^\la,q^\la{}_\m h^\m_a), \nonumber
\een
where $\si_0$ is the center of the quotient $GL_4/L$. 

Let $\Si_h$, coordinatized by $\wt\si^\m_a$, be the quotient 
of the bundle $Q$ by the kernel $\Ker_h\rho_h$ of the morphism
(\ref{a11}) with respect to the section $h$. This is isomorphic to the
bundle $\Si$ provided with the Lorentz structure of an
$L^hX$-associated bundle. Then the representation 
(\ref{a8}), which is constant on $\Ker_h\rho_h$, reduces to 
the representation
\ben
&& (\Si_h\times T^*X)\op\ot_{\Si_h} (\Si_h\times S^h) \to (\Si_h\times
S^h), \nonumber \\
&&(\wt\si,t^*)\mapsto \dot x_\la \wt\si^\la_a\g^a. \label{a12}
\een
Thence, one can think of a section $\wt h \neq h$ of the bundle $\Si_h$ 
as being an effective tetrad field, and can treat   
$\wt g^{\m\nu}=\wt h^\m_a \wt h^\nu_b\eta_{ab}$ as an effective metric.
A section $\wt h$ is not a true tetrad field, while 
$\wt g$ is not a true metric. Covectors
$\wt h^a=\wt h^a_\m dx^\m$ have the same representation by $\g$-matrices 
as the covectors $h^a=h^a_\m dx^\m$, while Greek indices go down and go
up by means of the background metric 
$g^{\m\nu}=h^\m_a h^\nu_b\eta_{ab}$.

Given a general covariant transformation $\wt f=\Phi\circ \wt f_h$ 
of the frame bundle $LX$, let us consider the morphism 
\beq
\wt f_Q:\quad Q\to \ol \Phi\circ \wt f_h(Q), \quad S^h \to \wt f_s(S^h), 
\quad T^*X\to \wt f (T^*X). \label{a9}
\eeq
This preserves the representation (\ref{a8}), i.e., 
$\g_Q\circ \wt f_Q =\wt f_s\circ \g_Q$, and yields the general
covariant transformation $\wt\si^\la_a\mapsto \dr_\m f^\la\wt\si^\m_a$
of the bundle $\Si_h$. 

Thus, we recover RTG \cite{log} in the case of a background tetrad
field $h$ and dynamic gravitational fields $q$.

\section{Gauge theory of RTG}

We follow the geometric formulation of field theory where a
configuration space of fields, represented by sections of a bundle
$Y\to X$, is the finite dimensional jet manifold $J^1Y$ of $Y$,
coordinatized by 
$(x^\la, y^i,y^i_\la)$ \cite{book,sar95}.  Recall that $J^1Y$ comprises
the equivalence classes of sections $s$ of $Y\to X$ which are
identified by their values and values of their first derivatives at points
$x\in X$, i.e.,
\be
y^i\circ s= s^i(x), \qquad y^i_\la\circ s= \dr_\la s^i(x).
\ee
A Lagrangian on $J^1Y$ is defined to be a horizontal density
\be
L=\cL(x^\la,y^i,y^i_\la)\om, \qquad \om =dx^1\cdots dx^n, \qquad n=\dim X.
\ee
The notation $\pi^\la_i=\dr^\la_i\cL$ will be used. 

A connection $\G$ on the bundle  $Y\to X$ is defined as a section of the jet
bundle $J^1Y\to Y$, and is given by the tangent-valued form 
\be
\G=dx^\la\ot(\dr_\la +\G^i_\la\dr_i).
\ee
For instance, a linear connection $K$ on the tangent bundle $TX$ reads
\be
K=dx^\la\ot(\dr_\la + K_\la{}^\al{}_\nu\dot x^\nu\frac{\dr}{\dr \dot x^\al}).
\ee
Every world connection $K$ yields the spinor connection  
\beq
K_h=dx^\la\ot[\dr_\la +\frac14 (\eta^{kb}h^a_\m-\eta^{ka}h^b_\m)(\dr_\la
h^\m_k - h^\nu_k K_\la{}^\m{}_\nu)L_{ab}{}^A{}_B y^B\dr_A]  \label{b3212}
\eeq
on the spinor bundle $S^h$ with coordinates $(x^\la,y^A)$, 
where $L_{ab}$ are generators (\ref{b3213}) \cite{sar97,book,ar}.

Using the connection $K_h$ and the representation $\g_Q$ (\ref{a8}),
one can construct the following Dirac operator on the product $Q\times S^h$:
\beq
\cD_Q= q^\la{}_\m h^\m_a\g^a D_\la, \label{a13}
\eeq
where $D_\la=y^A_\la - K^A_\la$ are the covariant derivatives
relative to the connection (\ref{b3212}). The operator (\ref{a13})
restricted to $q_0(X)$ recovers the familiar Dirac operator on $S^h$ 
for fermion fields in the presence of the background tetrad field $h$
and the world connection $K$. 

Thus, we obtain the metric-affine generalization of RTG where dynamic
variables are tensor gravitational fields $q$, general linear connections
$K$ and Dirac fermion fields in the presence of a background tetrad
field $h$ \cite{protv}. 
The configuration space of this model is the jet manifold 
$J^1Y$ of the product
\beq
Y=Q\op\times_X C_K\op\times_X S^h,\label{055}
\eeq
where $C_K=J^1LX/GL_4$ is the bundle whose sections are world connections $K$. 
The bundle (\ref{055}) is coordinatized by 
$(x^\m,q^\m{}_\nu, k_\al{}^\m{}_\nu,y^A)$. A total Lagrangian on this
configuration space is the sum 
\beq
L=L_{MA} +L_q(q,g) +L_D  \label{040}
\eeq
where $L_{MA}$ is a metric-affine Lagrangian, expressed into the curvature
\be
R_{\la\m}{}^\al{}_\bt = k_{\la\m}{}^\al{}_\bt - 
k_{\m\la}{}^\al{}_\bt+ k_\m{}^\al{}_\ve 
k_\la{}^\ve{}_\bt-k_\la{}^\al{}_\ve k_\m{}^\ve{}_\bt
\ee
and the effective metric $\wt\si^{\m\nu}=\wt\si^\m_a\wt\si^\nu_b\eta^{ab}$, 
the Lagrangian $L_q$ depends on tensor gravitational fields $q$ and the
background metric $g$, and 
Последний имеет вид
\be
&& \cL_D=\{\frac{i}{2}\wt\si^\la_q[y^+_A(\g^0\g^q)^A{}_B(y^B_\la-
 \frac14(\eta^{kb}\wt\si^{-1}{}_\m^a
-\eta^{ka}\wt\si^{-1}{}_\m^b)(\wt\si^\m_{\la k} -\wt\si^\nu_k
k_\la{}^\m{}_\nu)L_{ab}{}^B{}_Cy^C)- \\
&& \quad (y^+_{\la A} -
\frac14(\eta^{kb}\wt\si^{-1}{}_\m^a
-\eta^{ka}\wt\si^{-1}{}_\m^b)(\wt\si^\m_{\la k} -\wt\si^\nu_k
k_\la{}^\m{}_\nu)y^+_C L^+_{ab}{}^C{}_A(\g^0\g^q)^A{}_By^B]- \\ 
&&\qquad  my^+_A(\g^0)^A{}_By^B\}\nm{\wt\si}^{-1/2}, 
\qquad \wt\si=\det(\wt\si^{\m\n})
\ee
is the Lagrangian of fermion fields in
metric-affine gravitation theory [12-14], 
where tetrad fields are replaced with the 
effective tetrad fields $\wt \si$.
If
\beq
L_{AM}=(-\la_1R+\la_2)\nm{\wt\si}^{-1/2},  \quad
L_q=\la_3g_{\m\nu}\wt\si^{\m\nu}\nm{\si}^{-1/2}, \quad L_D=0, \label{039}
\eeq
where $R=\wt \si^{\m\nu}R^\al{}_{\m\al\nu}$, the familiar Lagrangian of
RTG is recovered.

\section{Energy-momentum conservation law}

We follow the standard procedure of constructing Lagrangian
conservation laws which is based on the first variational formula
\cite{book,jmp97}. This formula provides the canonical splitting
of the Lie derivative of a Lagrangian $L$ along a vector field $u$ on
$Y\to X$. We have
\ben
&& \dr_\la u^\la\cL +[u^\la\dr_\la+u^i\dr_i 
+(d_\la u^i -y^i_\m\dr_\la u^\m)\dr^\la_i]\cL = \label{bC30}\\
&& \qquad   (u^i-y^i_\m u^\m )(\dr_i-d_\la \dr^\la_i)\cL 
- d_\la[\pi^\la_i(u^\m y^i_\m -u^i) -u^\la\cL]. \nonumber
\een
This identity restricted to the shell
\be
(\dr_i- d_\la\dr_i^\la)\cL=0, \qquad d_\la=\dr_\la +y^i_\la\dr_i
+y^i_{\la\m}\dr_i^\m,
\ee
comes to the weak equality 
\ben
&& \dr_\la u^\la\cL +[u^\la\dr_\la+u^i\dr_i 
+(d_\la u^i -y^i_\m\dr_\la u^\m)\dr^\la_i]\cL \ap \label{J4}\\
&& \qquad -d_\la[\pi^\la_i(u^\m y^i_\m -u^i)-u^\la\cL]. \nonumber
\een
If the Lie derivative of $L$ along $u$ vanishes (i.e., $L$ is invariant
under the local 1-parameter group of gauge transformations generated by
$u$), we obtain the weak conservation law 
\be
0\ap - d_\la[\pi^\la_i(u^\m y^i_\m-u^i )-u^\la\cL].  
\ee

For the sake of simplicity, let us replace the bundle $Q$ in the product
(\ref{055}) with the bundle $\Si_h$, and denote 
\be
Y'=\Si_h\op\times_XC_K\op\times_X S^h.
\ee
There exists the canonical lift on $Y'$ of every vector field 
$\tau$ on $X$:
\ben
&&\wt\tau = \tau^\m\dr_\m +
(\dr_\nu\tau^\al k_\m{}^\nu{}_\bt - \dr_\bt\tau^\nu
k_\m{}^\al{}_\nu - \dr_\m\tau^\nu
k_\nu{}^\al{}_\bt +\dr_{\m\bt}\tau^\al)\frac{\dr}{\dr k_\m{}^\al{}_\bt}+
 \dr_\nu\tau^\m \wt\si^\nu{}_a  \frac{\dr}{\dr \wt\si^\m_a}
+\label{051} \\ 
&& \quad \frac14 (\eta^{kb}h^a_\m-\eta^{ka}h^b_\m) (\tau^\la\dr_\la
h^\m_k - h^\nu_k\dr_\nu\tau^\m) (L_{ab}{}^A{}_B y^B\dr_A
-L_{ab}{}^c{}_d\wt\si^\nu_d
\frac{\dr}{\dr \wt\si^\nu_c}), \nonumber
\een
where $L_{ab}{}^c{}_d$ are generators (\ref{b3278}). This lift is the
generator of gauge transformations of the bundle $Y'$ induced by morphisms
$\wt f_Q$ (\ref{a9}). Its part acting on Greek indices is the familiar
generator of general covariant transformations, whereas that acting on
the Latin ones is a local generator of vertical Lorentz gauge transformations.

Let us examine the weak equality (\ref{J4}) in the case of the
Lagrangian (\ref{040}) and the vector field (\ref{051}). In contrast
with $L_q$, the Lagrangians $L_{MA}$ and $L_D$ are invariant under the
above-mentioned gauge transformations. Then, using the results of
\cite{sar97,book}, we bring (\ref{J4}) into the form
\ben
&& \dr_\la(\tau^\la\cL_q) + (\dr_\al\tau^\m \wt g^{\al\nu}
+\dr_\al\tau^\nu\wt g^{\al\m})\frac{\dr\cL_q}{\dr\wt g^{\m\nu}} \ap
\label{d20} \\
&& \qquad  d_\la(2\tau^\m \wt g^{\la\al}\frac{\dr\cL_q}{\dr\wt g^{\al\m}} +
\tau^\la\cL_q - d_\m U^{\m\la}), \nonumber
\een
where
\be
U= 2\frac{\dr\cL_{AM}}{\dr 
R_{\m\la}{}^\al{}_\nu}(\dr_\nu\tau^\al-
k_\si{}^\al{}_\nu\tau^\si)
\ee
is the generalized Komar superpotential of the energy-momentum of
metric-affine gravity \cite{sar97,book,giach}. 

A glance at the expression (\ref{d20}) shows that, if the Lagrangian
$L$ (\ref{040}) containes the Higgs term $L_q$, the energy-momentum
flow is not reduced to a superpotential, and the familiar covariant
conservation law  
\beq
\wt\nabla_\al t^\al_\la \ap 0, \qquad t^\al_\la=2\wt g^{\al\m}
\frac{\dr\cL_q}{\dr \wt g^{\m\nu}}, \label{061}
\eeq
takes place. Here, 
$\wt\nabla_\al$ are covariant derivatives relative to the
Levi--Civita connection of the effective metric $\wt g$.  
In the case of the standard Lagrangian $L_q$ (\ref{039}) of RTG, the 
equality (\ref{061}) comes to the well-known condition 
\be
\nabla_\al (\wt g^{\al\m}\sqrt{\nm{\wt g}})\ap 0,
\ee 
where $\nabla_\al$ are covariant derivatives relative to the
Levi-Civita connection of the background metric $g$. On solutions
satisfying this condition, the energy-momentum flow in RTG reduces to
the generalized Komar superpotential just as it takes place in General
Relativity \cite{nov}, Palatini formalism \cite{nov93,bor},
metric-affine and gauge gravitation theories
\cite{sar97,book,giach}.

\end{document}